\documentclass[openacc]{rstransa}

\newtheorem{theorem}{\bf Theorem}[section]
\newtheorem{condition}{\bf Condition}[section]

\titlehead{Research}

\begin{document}

\title{Similarities and differences between solar and stellar flare pulsation processes}

\author{
F. Reale$^{1,2}$}

\address{$^{1}$Department of Physics and Chemistry, University of Palermo, via Archirafi 36, 90134 Palermo, Italy\\
$^{2}$INAF - Osservatorio Astronomico di Palermo, Piazza Parlamento 1, 90134 Palermo, Italy}

\subject{Astrophysics}


\corres{Fabio Reale\\
\email{fabio.reale@unipa.it}}

\begin{abstract}
Quasi-periodic pulsations (QPPs) are oscillatory signatures commonly detected in the light curves of solar and stellar flares, offering valuable diagnostics of the underlying magnetic and plasma processes. This review compares the observational characteristics, detection methods, and physical interpretations of QPPs in both solar and stellar contexts. Solar flare QPPs, extensively studied in X-rays and EUV bands using instruments such as GOES, STIX, and Fermi, display typical periods of tens of seconds and show correlations with flare duration and magnetic loop length. Stellar QPPs, observed in X-rays and white light by missions such as Kepler, TESS, and XMM-Newton, exhibit much longer periods - ranging from minutes to hours - consistent with larger-scale magnetic structures 
in more active stars. Despite differences in scale and observing band, statistical and comparative studies reveal common scaling relations and damping behaviors, suggesting that both solar and stellar QPPs are manifestations of the same fundamental mechanisms, likely magnetohydrodynamic oscillations or oscillatory reconnection within flare loops. The comparison underscores a continuity between solar and stellar magnetic activity, linking the  {solar detailed physical processes} to stellar-scale phenomena and providing constraints for future models and surveys.
\end{abstract}


\begin{fmtext}

\end{fmtext}


\maketitle

\section{Introduction}

Solar flares are a manifestation of a sudden release of magnetic energy into heat, {particle acceleration and so on,} due a rapid rearrangement of a strong stressed magnetic field. The Sun shows that the energy release is highly impulsive and localized. Flares are visible as brightening peaks in all energy bands, from radio to gamma rays, but they are traditionally best detected in the high energy bands. First resolved observations were performed in the soft X-rays \cite{Vaiana1973a} (see \cite{Testa2023} for a review). Their classification (C to X) is based on the flux peak as observed in the X-rays 1-8 \AA\ band by the Geostationary Operational Environmental Satellite (GOES) series. All flares involve the rearrangement of magnetic structures in the solar corona, and {generally detected} in the X-rays. As such the X-ray band is the reference band for solar flares. 
Very intense flares produce radiation also at lower energies, and make bright layers below the corona, deeper in the chromosphere and photosphere. {These are generally more difficult to detect and, therefore, to apply systematic and statistical analysis to them.}
The light curve of many solar flares shows an interesting modulation that has often an almost sinusoidal trend. The period and the amplitude can somewhat change during the evolution and therefore these are \textit{Quasi-Periodic Pulsations} (QPPs). 

On the other hand, flares are also commonly observed on stars in the soft X-ray band, and usually share observational features with solar flares, such as the short impulsive phase followed by the slower decay, temperatures well above those measured outside of the flare  \cite{Haisch1983}. As such, it is commonly accepted that stellar flares have the same origin and the same physical mechanisms as solar flares, i.e., they are driven by local rearrangements of stressed coronal magnetic fields. Along this line, they have, for instance, been interpreted and described with solar loop hydrodynamic models \cite{Reale1988a,Reale2004a,Favata2005a,Schmitt2008a,Reale2018a}. 

Much more extensive surveys of stellar flares have been conducted recently in the white light band in the framework of surveys of exoplanets by the Kepler \cite{Borucki2010a} and the Transiting Exoplanet Survey Satellite (TESS) \cite{Ricker2015a} missions. Therefore, large databases are available in this band, and it is possible to perform statistical studies with results at high statistical significance.

However, it is not trivial to find a way to compare systematic surveys of solar flares in the high energy bands with surveys of stellar flares in the white light band. In the following, we will assume that we are mostly observing the same phenomenon, in a different band simply because stellar flares are very strong and {the released energy propagates} effectively from the corona downward to the chromosphere and photosphere, as it occurs on a smaller scale in solar two-ribbon flares. This will allow us to make a comparison between rather large populations and therefore between robust statistical quantities.

Whereas flares, although highly localized, can be rather chaotic events, which can involve several magnetic structures altogether or progressively, even with fine structuring \cite{Doschek2005a,Testa2020a}, QPPs are instead an indication of a coherent evolution of a single perturbation and in a single perturbed system. They become then a valuable probe of the physical and geometrical conditions where we have no spatial resolution as in other stars. It is also interesting to investigate the possible drivers that overlap the flare mechanisms or might shed more light on them (some reviews \cite{Nakariakov2009a,McLaughlin2018a,Wang2021a}).

Exhaustive and extensive reviews which discuss QPPs both in solar and stellar flares are available \cite{VanDoorsselaere2016a,McLaughlin2018a,Nakariakov2019a,Kupriyanova2020a,Zimovets2021a}.
{This is an updated review where} I provide my personal view about the comparison of pulsations observed in solar flares to those in stellar flares, in particular focussing on general properties derived from systematic analyses.



\section{Solar and stellar QPPs}
\label{sec:qpp}

\subsection{Solar QPPs}
\label{sec:solar}

As a first step, it is worthwhile to briefly list the main techniques used to detect QPPs in solar flares and measure their periods and other features. Several analytical methodologies are available to search for stable periodic signals in various forms of flare emission.
QPPs are typically identified by searching for significant peaks in the power spectrum of flare lightcurves using Fourier analysis \cite{Inglis2016a,Lim2025a}  or other periodogram-based approaches \cite{Pugh2017a,Pugh2017b,Pugh2019a}  For instance, the Lomb-Scargle method is a specific periodogram-based technique widely employed, with some studies indicating that 70\% of examined flares showed statistically significant oscillations using this method, which were then confirmed by the autocorrelation method for 48\% of flares (see below) \cite{Szaforz2025a}. The Automated Flare Inference of
Oscillations (AFINO) methodology \cite{Inglis2016a} has been applied to GOES lightcurves \cite{Hayes2020a}  to detect a distinct class of stationary or weakly nonstationary QPPs that exhibit significantly stable periodic signals throughout the flare. Fourier analysis has also been used to identify QPPs in integrated light curves of EUV brightenings. Another powerful method is wavelet transforms. This method is sometimes applied to the time derivative of GOES channels rather than detrended signals. The presence of QPPs is defined by identifying power in the wavelet power spectrum above a red-noise model at a high confidence level, such as 99.7\%  \cite{Simoes2015a}. Recently, also machine-learning techniques, e.g., Fully Convolution Network (FCN), have been tested, providing results comparable to the other established methods \cite{Belov2024a}.

\begin{figure}[!h]
\centering\includegraphics[width=5in]{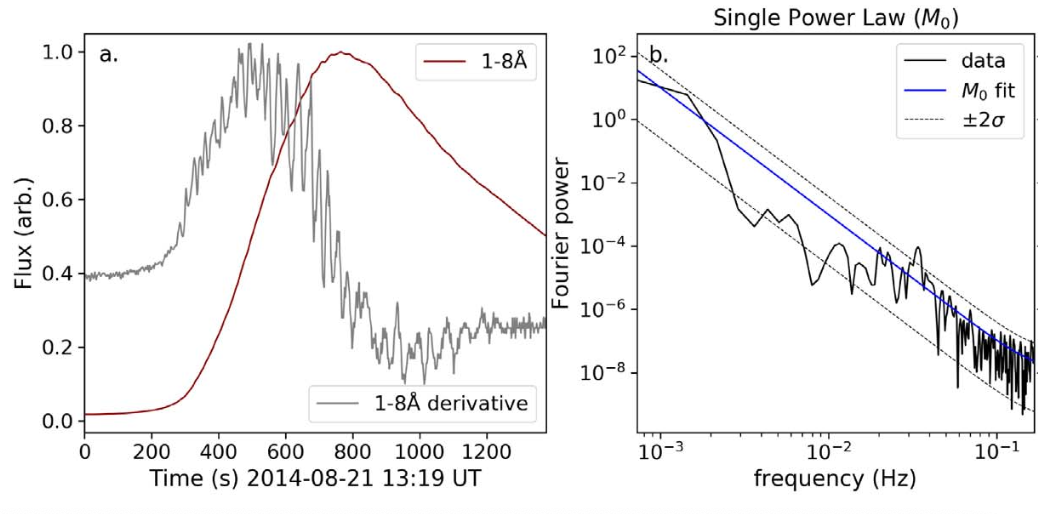}
\caption{Flare light curve showing quasi-periodic pulsations. Examples of AFINO analysis applied to a solar M3.4 flare. The left panel (a) shows the 1–8 \AA\ lightcurve (red) and its associated time derivative (gray). A corresponding Fourier power spectrum
of the red lightcurve is plotted (black) in the right panel (b), together with a model fit  (blue). The dashed gray lines mark the 2.5\% and 97.5\% quantiles relative to the power-law component of each model to help illustrate enhanced peaks in the spectra. (From \cite{Hayes2020a}).}
\label{fig:hayes0}
\end{figure}

These methods are applied to lightcurves obtained from various instruments and across the electromagnetic spectrum, including radio to gamma-ray ranges, in both non-thermal and thermal radiation. For example, studies have analyzed GOES X-ray lightcurves (e.g., 1–8 \AA), hard X-ray (HXR) observations, and extreme-ultraviolet (EUV) brightenings. More recently, STIX quick-look light curves have been used, with the strongest maximum in Lomb-Scargle periodograms often observed in the 4–10 keV energy channel, which is dominated by thermal emission \cite{Szaforz2025a}.

\begin{figure}
    \centering
    \includegraphics[width=4in]{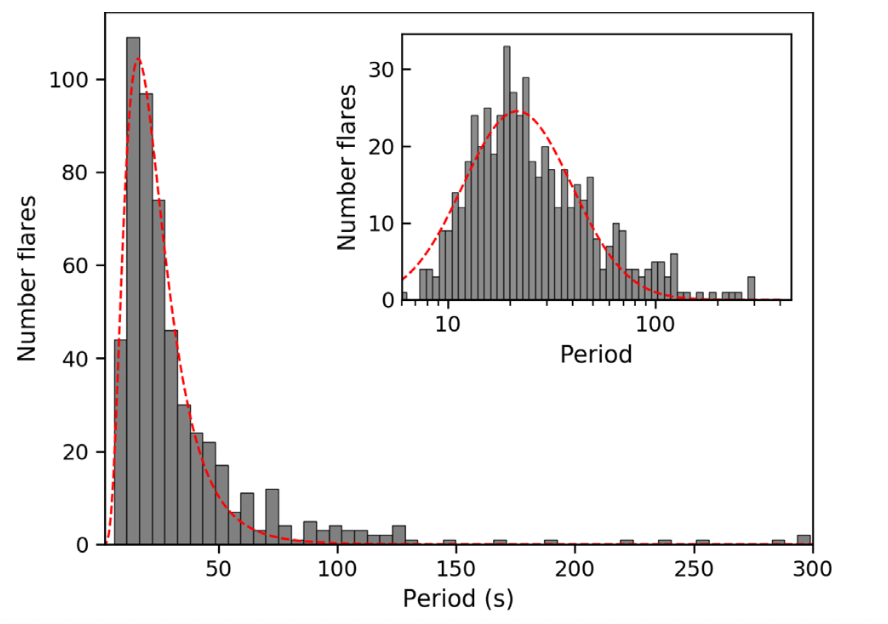}
    \caption{Histogram of identified QPP periods from the statistical survey of X-, M-, and C-class solar flares. The  best-fit log-normal distribution (inset) has a mean period of 21.6s (red dashed line). (From \cite{Hayes2020a}).}
    \label{fig:Hayes20_1}
\end{figure}

\begin{figure}
    \centering
    \includegraphics[width=5.in]{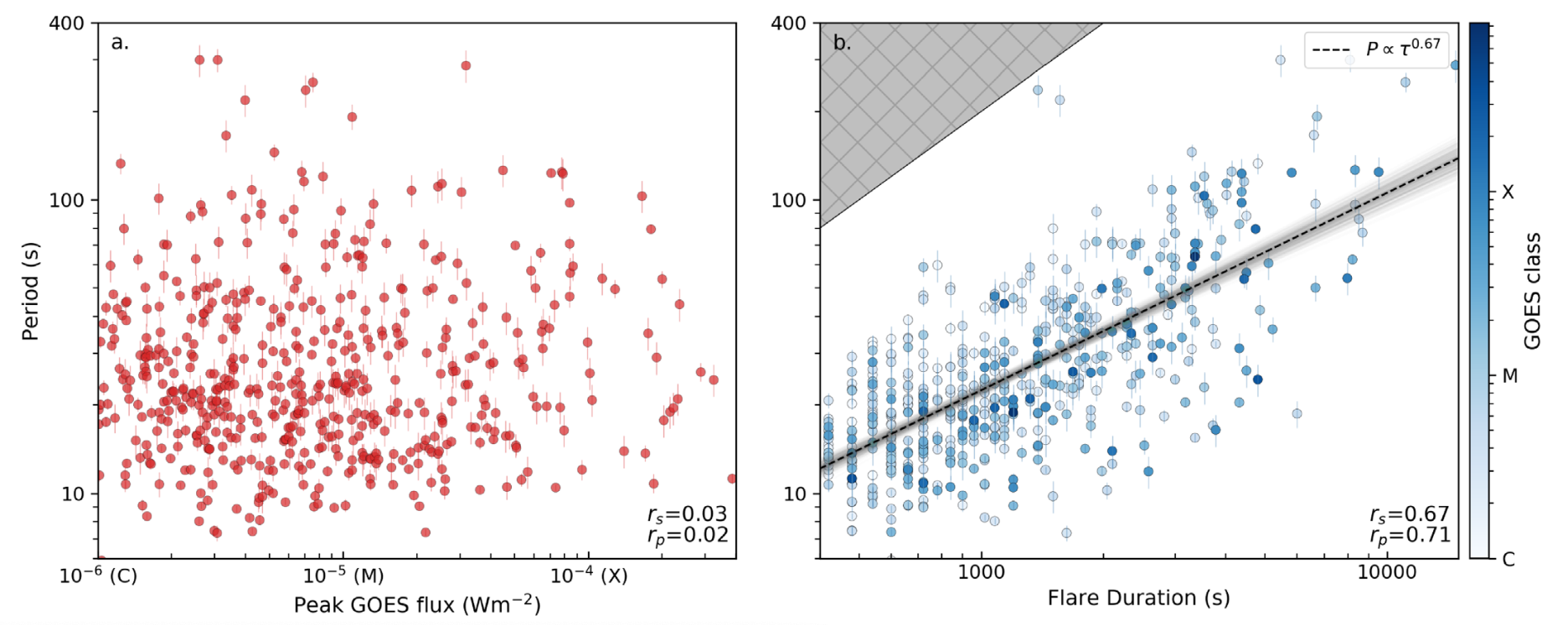}
    \caption{Scatter plots of QPP periods with (a) flare size and (b) flare duration.  In the hatched region duration–period combinations cannot be detected. A power-law fit is shown in panel b) (dashed line).
The correlation coefficients are labeled in the
bottom right of both scatter plots. (From \cite{Hayes2020a})}
    \label{fig:hayes20_2}
\end{figure}

To assess general feature of solar QPPs it is important to have a large and homogeneous baseline for statistics. In this sense, though with no spatial resolution, the continuous monitoring  by the GOES satellite series in the soft X-ray band provides an ideal source of data. Probably the most extensive systematic investigation of flare QPPs observed with GOES was performed by Hayes et al. \cite{Hayes2020a}, in the previous solar cycle. The selection was on relatively long events, longer than 400 s, and as intense as C-class or more. They were left with a large number of over 5000 events, up to X9.3. The GOES light curves were analysed with the  AFINO method \cite{Inglis2016a}, and periodograms were derived, searching for bumps in the Fourier power spectrum in the range 6 s to 300 s (Fig.\ref{fig:hayes0}). 
Interestingly, it was found that the fraction of flares showing QPPs increases with the intensity, from 7\% for C-class to 29\% for M-class and to 46\% for X-class. The most important parameter is the period, which was found {to be fit by} a log-normal distribution, with a mean of 22 s and {$1-\sigma$} range between 12 s and 40 s (Fig.\ref{fig:Hayes20_1}). No correlation was found between the period and the flare intensity (Fig.\ref{fig:hayes20_2}).

A significant correlation was found instead with the duration of the flare, similar to \cite{Pugh2017b,Pugh2019a} (Fig.\ref{fig:hayes20_2}):
\begin{equation}
    P \propto \tau^{0.7} 
\end{equation}
where P is the QPP period and $\tau$ is the duration of the flare.
Although GOES satellite has no spatial resolution, and therefore a direct connection with the size of the flaring region cannot be found from the X-ray data, a proxy for that is the search for correlation with the separation of the flare ribbons. Flare ribbons are bright parallel strips visible in the H$\alpha$ line, which mark the footpoints of flaring loop arcades. The more the distance between ribbons, the longer are presumable connecting loops. Of the about 2000 flares which were found with corresponding ribbon, about 10\% showed QPPs. Indeed a positive correlation for ribbon separation was found with the following scaling:

\begin{equation}
    P \propto d^{0.6} 
\end{equation}
where $d$ is the ribbon separation. This suggests that QPP periods are connected to the length of the magnetic tubes where they propagate.
It was also shown that there is no particular connection between the presence of QPPs and the ejection of plasma from the flare site (CMEs).
Also in the EUV band, empirical scaling laws were inferred between the QPP periods and the ribbon properties in flares observed with Atmospheric Imaging Assembly (AIA) 1600 \AA\ and Helioseismic and Magnetic Imager (HMI) on-board the Solar Dynamics Observatory, although with loose correlation \cite{Pugh2019a}.

An interesting issue to investigate whether QPPs are triggered by the same or by different mechanisms is to compare QPPs in the rise and those in the decay phase of flares. Only about 10\% of events were selected for this comparison, because the rise phase is fast and difficult to resolve.
A Kolmogorov-Smirnov (K-S) test showed that the distributions of the QPP periods belong to different populations.

Other studies also investigated QPPs observed with the GOES series. The wavelet analysis \cite{Torrence1998a}  was applied to a sample of 35 X-class events in Cycle 24 finding an average characteristic time scale between about 15 s and 60 s \cite{Simoes2015a}. 
Another previous analysis applied to a smaller flare sample, an interesting comparison was performed between flares observed with the GOES series and with the Fermi/Gamma-ray Burst Monitor (GBM) 15–25 keV X-ray data \cite{Inglis2016a}. Although they share a preferred characteristic QPP timescale of about 5–30 s, with no dependence on flare magnitude, it is suggested that overall there is
greater sensitivity to QPP-like signals in GOES data than in the higher energy 15–25 keV data obtained by Fermi/GBM. 
Moving to recent observations in the lower energy EUV band, only 3\% of over 22000 brightenings observed by the Solar Orbiter/Extreme Ultraviolet Imager were found with QPPs with Fourier analysis, with periods between 15 and 260 s \cite{Lim2025a}. 
At the other extreme, in the hard X-ray band,  35\% to 65\% of 74 flares observed with the HXT instrument onboard the Yohkoh satellite showed QPPs, at increasingly higher energies, with periods between 20 s to about 300 s \cite{Szaforz2019a}. More recently, in a sample of over 300 flares from B4 to X1 class, showing QPPs with the STIX instrument on-board Solar Orbiter, 129 showed well-defined pulsations with periods in the range of 40–1400 s \cite{Szaforz2025a}.

An interesting study concerns systematic analysis of flares from a single active region \cite{Pugh2017b}. No correlations were found between QPPs periods and the AR area, bipole separation distance, or average magnetic field strength, therefore indicating that the modulation depends only on local flare characteristics.
Another interesting property of QPPs, and in particular those detected in the decay phase, is that they are damped on the way. A systematic study of this effect was performed for comparison on both solar and stellar flares \cite{Cho2016a}.
In general, there is agreement that, in many cases, the QPP periods are correlated with the length of the involved flaring structures.

There are several candidate mechanisms invoked to drive solar  (and stellar) flare Quasi-Periodic Pulsations (QPPs). 
{A list of them includes include MHD wave modes or oscillations, self-oscillatory mechanisms, periodic and/or wave-driven reconnection, loops merging, MHD flow over-stability,  equivalent LRC-like oscillations, and thermal-dynamical cycles. Examples of wave modes are the fundamental magnetosonic harmonics of magnetic flux tubes, such as kink, sausage, fluting, torsional and acoustic modes \cite{McLaughlin2018a}.  According to \cite{Kupriyanova2020a}, we might generally classify the drivers into three main categories}:

\begin{enumerate}
    \item Direct modulation of radiation emission caused by magnetohydrodynamic (MHD) waves or various types of electromagnetic oscillations.
    \item Modulation of the efficiency of the energy release process
    \item Spontaneous, quasi-periodic energy release
\end{enumerate}

{We mention also a model based on the Kelvin-Helmholtz instability \cite{Fang_2016}.}
More specifically, several leading ideas and observations point to particular drivers. The periods of QPPs overlap with the expected timescales of MHD wave modes in the solar corona, which suggests a relation to resonant MHD wave processes occurring at the flaring site \cite{Hayes2020a}.
Oscillatory magnetic reconnection is also considered a strong candidate for causing pulsations in flare emission \cite{Schiavo2024a}. Oscillations in the reconnection process can arise intrinsically or through modulation from some external driver.  One way oscillatory reconnection can occur is when an incoming fast wave pulse impinges on a magnetic X-point. Other models include a guide field, as customary of the coronal flux tubes, and allow reconnection to arise naturally from the initial magnetic field without an imposed external driving pulse. For instance, the merger of twisted magnetic flux ropes can lead to oscillatory reconnection, as well as the slow outward propagation of wave-like disturbances from the reconnection site after the merger \cite{Stewart2022a}.
Another possibility is suggested by the observed scaling law between QPP period and ribbon separation distance (where longer ribbon separation, acting as a proxy for loop length, hosts longer QPP periods) \cite{Pugh2019a}. This suggests that ongoing reconnection at increasingly higher altitudes, producing new, increasingly longer loops, could be related to QPPs {in a non-linear way}. This interpretation might also explain why QPP periods are often longer in the decay phase compared to the impulsive phase, as longer loops form in later stages of the flare, especially in eruptive flares associated with CMEs. {There is recent evidence of a strong correlation between QPPs with harmonics and magnetic reconnection in a multiwavelength observation of the rising phase of a two-ribbon solar flare \cite{Song2025} (see also Sec.\ref{sec:disc}).}
Finally, another scenario involves slow-mode MHD waves triggering episodes of magnetic reconnection at successive locations \cite{Nakariakov2011,Stewart2022a}. However, this might be difficult in the impulsive phase of a flare because the non-steady evolution of the magnetic field required for rapid energy release would tend to disrupt the identities of eigenmodes as the impulsive phase develops.

\subsection{Stellar QPPs}
\label{sec:stellar}

QPPs have a long history of detections in the light curves of stellar flares. 
As for solar flares, there are several methods for determining QPP periods in stellar flares. The Fourier Transform technique \cite{Ramsay2021a} can be applied in different ways, i.e., with detrending (smoothing the data) but without accounting for background colored noise, and without detrending, but specifically accounting for the background colored noise. The empirical Mode Decomposition (EMD) method efficiently decomposes the original time series into Intrinsic Mode Functions (IMFs) \cite{Cho2016a}. The IMF with the slowest characteristic timescale can be used as a trend for the original signal. EMD also includes a self-consistent detrending and an assessment of the statistical significance of the revealed intrinsic oscillatory modes against background colored noise. This method was used in analyzing QPPs in both solar and stellar X-ray flares. As for solar QPPs  (Section \ref{sec:qpp}\ref{sec:solar}), the wavelet analysis, can be applied to detect quasi-periodic signals in stellar flares \cite{Welsh2006a,Pandey2009a,Balona2015a,Lopez-Santiago2016a}. {Also Neural Networks (Fully Convolutional Networks - FCNs) \cite{Wang2016,Ismail-Fawaz2019a}} have been successfully used to detect QPPs in stellar flares. For example, an FCN was tested on 11 stellar flare light curves with strong evidence of decaying QPPs found in Kepler observations by C. E. Pugh et al. \cite{Pugh2016a}, {and it detected QPPs in 7 of these 11 flares \cite{Belov2024a}}. In this case,  the large-scale
Kepler flare catalogue included over 2000 events. This machine-learning technique resulted in a 7\% QPP detection rate with a probability above 95\%, in agreement with \cite{Ramsay2021a}.

In the soft X-ray band a limited survey of the presence of oscillations in flare light curves  was performed on the flares captured in a deep observation of the Orion star-forming region by the Chandra mission [Chandra Orion Ultradeep Project, COUP, \cite{Getman2005a,Favata2005a}. A wavelet analysis \cite{Torrence1998a} was applied to 5 stellar flares in COUP and showed periods between 2 and 10 ks \cite{Lopez-Santiago2016a}. 
About 40 QPPs were detected in light curves of stellar X-ray flares detected in the 0.3–2 keV band of XMM-Newton in a sample of 22 stars, mostly of M-type \cite{Cho2016a}. They are compared with analogous ones observed in a set of solar flares observed with RHESSI and found to have similar shapes, only differing in the much longer periods (about 20 min) and damping times (about 30 min).
{It is to be noted that stellar flares detected in the X-rays are usually much hotter than solar flares. Therefore, the spectrum of the thermal emission is quite harder than that of solar flares and extends to higher energies, and this makes the comparison of solar and stellar flares less obvious in similar bands. It is also much more difficult to distinguish thermal and non-thermal emission, the stellar flare spectra are generally interpreted in terms of one or more thermal components in the X-ray bands of XMM-Newton and Chandra telescopes. So there is no specific detection of QPPs in stellar flares for non-thermal emission at high energies.}

\begin{figure}[!h]
\centering\includegraphics[width=5in]{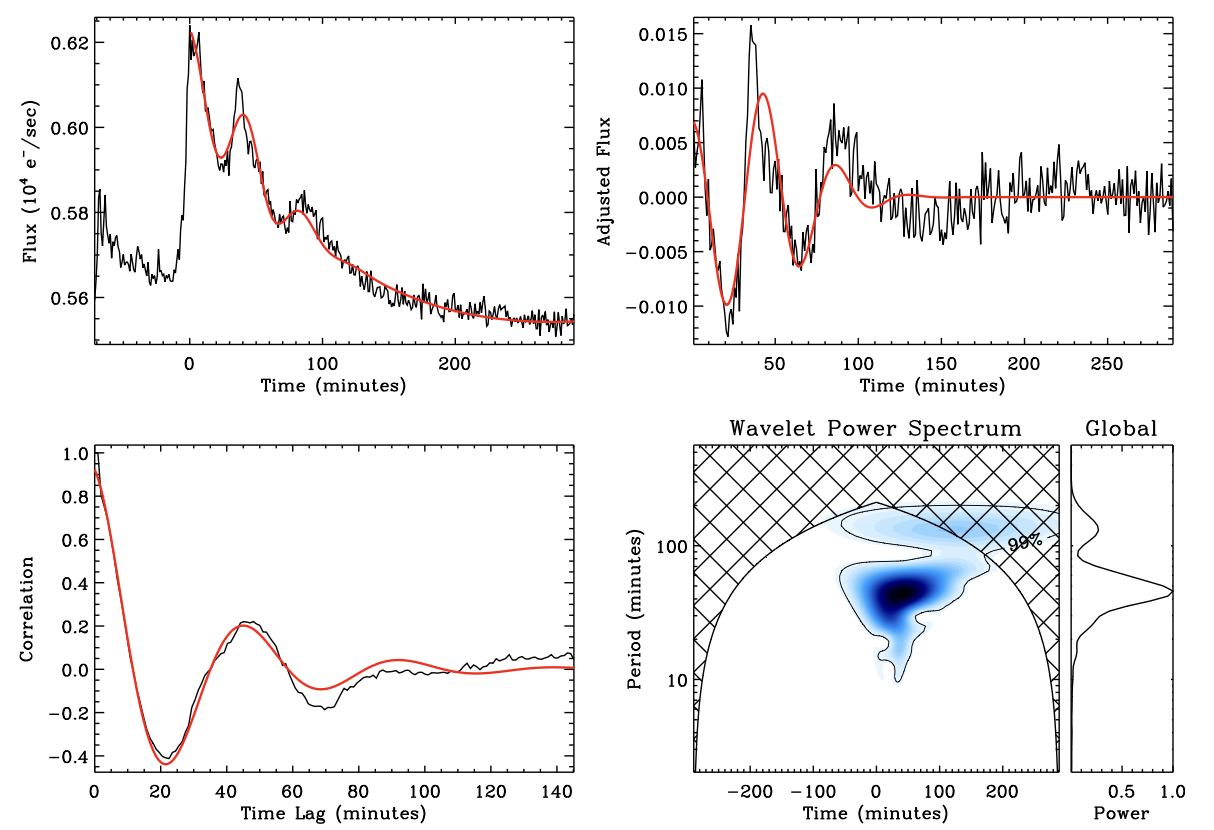}
\caption{Top left: light curve of a stellar flare by Kepler and 
the result of a least-squares fit to the flare decay combined with the QPPs (red  line). Top right: light curve after subtracting the flare decay trend with a decaying sinusoidal fit ( red line). Bottom left: 
autocorrelation function, with a fitted exponentially decaying sinusoid (red). Bottom right: wavelet
spectrum of the top right plot and the global wavelet spectrum.(from \cite{Pugh2016a}).}
\label{fig:pugh16_0}
\end{figure}

As mentioned above, white-light emission in solar flares is difficult to detect, and even the strongest solar flares are barely detected in space
observations of total solar irradiance.  The Sun would not be detected as a flare star by the Kepler mission.
However, there is a strong correspondence between white-light
emission and hard X-ray emission  \cite{Hudson2006a}.
Kepler \cite{Borucki2010a} made white-light observations of around
150000 sources between 2009 and 2013. The mission was mainly devoted to the detection of extrasolar planets (exoplanets), but its extensive surveys allowed to detect a wealth of stellar flares. 
QPPs were in fact detected in the decay phase of white-light stellar flares observed by Kepler.
An initial study addressed very strong flares (superflares) in the Kepler survey \cite{Balona2015a}. 18\% of the analyzed 250 flares, with typical duration of 1 hour, showed modulation in the decay of the light curve. The periods are between 5 and 100 min. 
No correlation between these time-scales and any stellar parameter was found.  Damped oscillations after flare maximum were found in a few flares, with periods between 5 and 15 min.
A more systematic study was performed later \cite{Pugh2016a}.
Out of the about 1500 flares detected on about 200 different stars, about 60 show QPPs in the Kepler light curve. Evidence of decaying oscillations is shown by 11 of them (Fig.\ref{fig:pugh16_0}). The host stars range from F- to M-type.

\begin{figure}[!h]
\centering\includegraphics[width=5in]{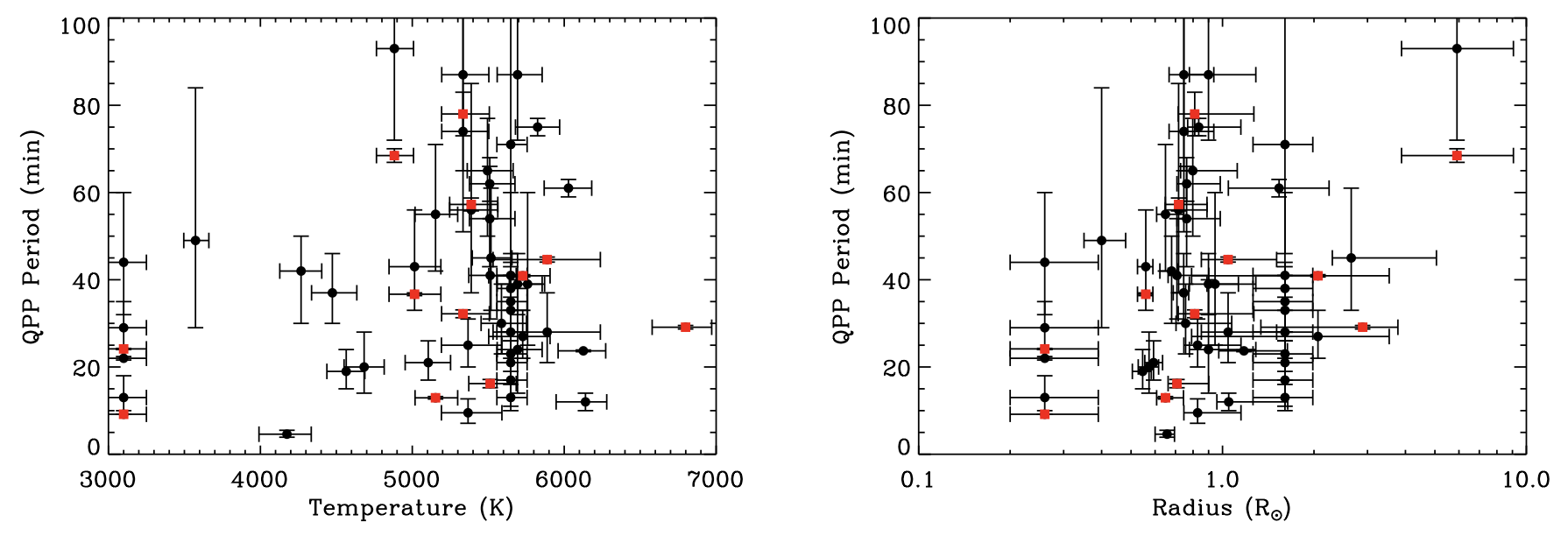}
\caption{ Left: scatter plot of stellar effective temperature and QPP period. Right: scatter plot of stellar radius and QPP period. (from \cite{Pugh2016a}}
\label{fig:pugh16a}
\end{figure}

Searching for correlations of QPPs with flare and global stellar parameters, a first but very important "negative" result is that, also in this more extensive survey, none of the correlation coefficients show a clear relationship between the QPP period and the global stellar parameters, nor with the total flare energy (Fig.\ref{fig:pugh16a}). This confirms that probably the pulsations are related to the local flaring system only.
Periods are measured generally in the range between 10 and 100 min, but this is limited by the data cadence and the flare durations.
A clear oscillation was detected also in a flare of Kepler K2 campaign in one star of the Pleiades cluster, with a period of about 10 min \cite{Guarcello2019a}. {In this work, the emission in the hardest part of the detected X-ray spectrum (3 - 7.9 keV) is discussed, and indeed a hint of oscillation might be present, but it is not possible to assess whether it is thermal or non-thermal emission.}

The TESS mission was launched in 2018, has made photometry of a large fraction of the sky in the white light band. It has observed many flares on solar-type stars \cite{DoyleL2020a} and M-dwarfs \cite{Ramsay2020a}.
Of the total sample of about 500 flaring stars, the light curves of about 1/3 were analyzed in detail \cite{Ramsay2021a}. A combination of detection methods with Fourier and empirical mode decomposition was used \cite{Broomhall2019a}. A number of 11 QPPs were detected in the decay of flares of 7 stars only.  The measured periods are between 10 and 70 min.
An exception is represented by the oscillations detected in flares on dMe stars in the UV band (GALEX satellite) \cite{Welsh2006a}, with periods around 1 min.

The underlying mechanisms responsible for stellar QPPs are probably similar to those for solar flares {(see Sec.\ref{sec:qpp}\ref{sec:solar})}. One shared mechanism could be, for instance, the natural magnetoacoustic oscillations of the flaring or adjacent coronal loops.
Specifically, proposed MHD mechanisms for stellar QPPs also include the resonant absorption (mode coupling) \cite{Pandey2009a,Pugh2016a}. {This results into a damping mechanism for kink waves, where the damping time is proportional to the period. Thermal conduction \cite{Lopez-Santiago2016a} is another possible damping mechanism, although it has been shown that misbalance between heating and cooling processes might be more efficient \cite{Kolotkov2019}}.
{Among the mechanisms listed in Sec.\ref{sec:qpp}\ref{sec:solar}, sausage modes have been specifically considered to be detectable in the light curve of flaring stars, and this allows to estimate the magnetic field strength at the top of the loop from the oscillation's amplitude
\cite{Balona2015a,Kupriyanova2020a}.} 
Regarding kink modes, they are not supposed to cause changes in density and are therefore generally not expected to be detected in the light curve of flaring stars \cite{Ramsay2021a}.
Other alternative proposed scenarios for stellar QPPs include slow magneto-acoustic modes, fast modes, and Kelvin-Helmholtz instabilities \cite{Lopez-Santiago2016a,Ramsay2021a}. However, many of these processes are thought to have negligible effects on the soft X-ray light curve.



\section{Discussion on similarities and differences between Solar and Stellar QPPs}
\label{sec:disc}

It is not trivial to perform systematic analysis of QPPs in stellar flares, and therefore to derive general properties. QPPs are best and more commonly observed in the X-ray band. In this band it is difficult to observe large samples of stellar flares. They remain serendipitous events, because they can only be obtained from X-ray telescopes in space missions, which share the observing time with many other possible targets, from supernova remnants (SNR), to compact binaries, to clusters of galaxies. On the other hand, stellar flares are often much more powerful than solar ones, and can be detected in the visible band, whereas they would be not on the Sun as a star. Therefore, flare observations can be obtained more easily from photometric surveys of missions entirely devoted to the detection of extrasolar planets in the white light band. However, this is not the X-ray band and the very occurrence of pulsations in this band is obviously much less frequent. It is already quite a surprising circumstance that there are QPP detections in this band, whereas none is detected on the Sun, and this is entirely due to the much higher intensity of stellar events.

The similarities and differences between QPPs in solar and stellar flares have been thoroughly discussed in other extensive reviews \cite{Nakariakov2009a,McLaughlin2018a,Wang2021a,VanDoorsselaere2016a,Nakariakov2019a,Kupriyanova2020a,Zimovets2021a}. It is to be remarked that solar and stellar QPPs share many features, which suggest that they have a common origin. However, the overall scale is different.
Here we will concentrate on the most commonly found values of the main parameter of the QPPs, which is the period, as derived from the most extensive surveys of solar and stellar flares.

\begin{figure}[!h]
\centering\includegraphics[width=2.5in]{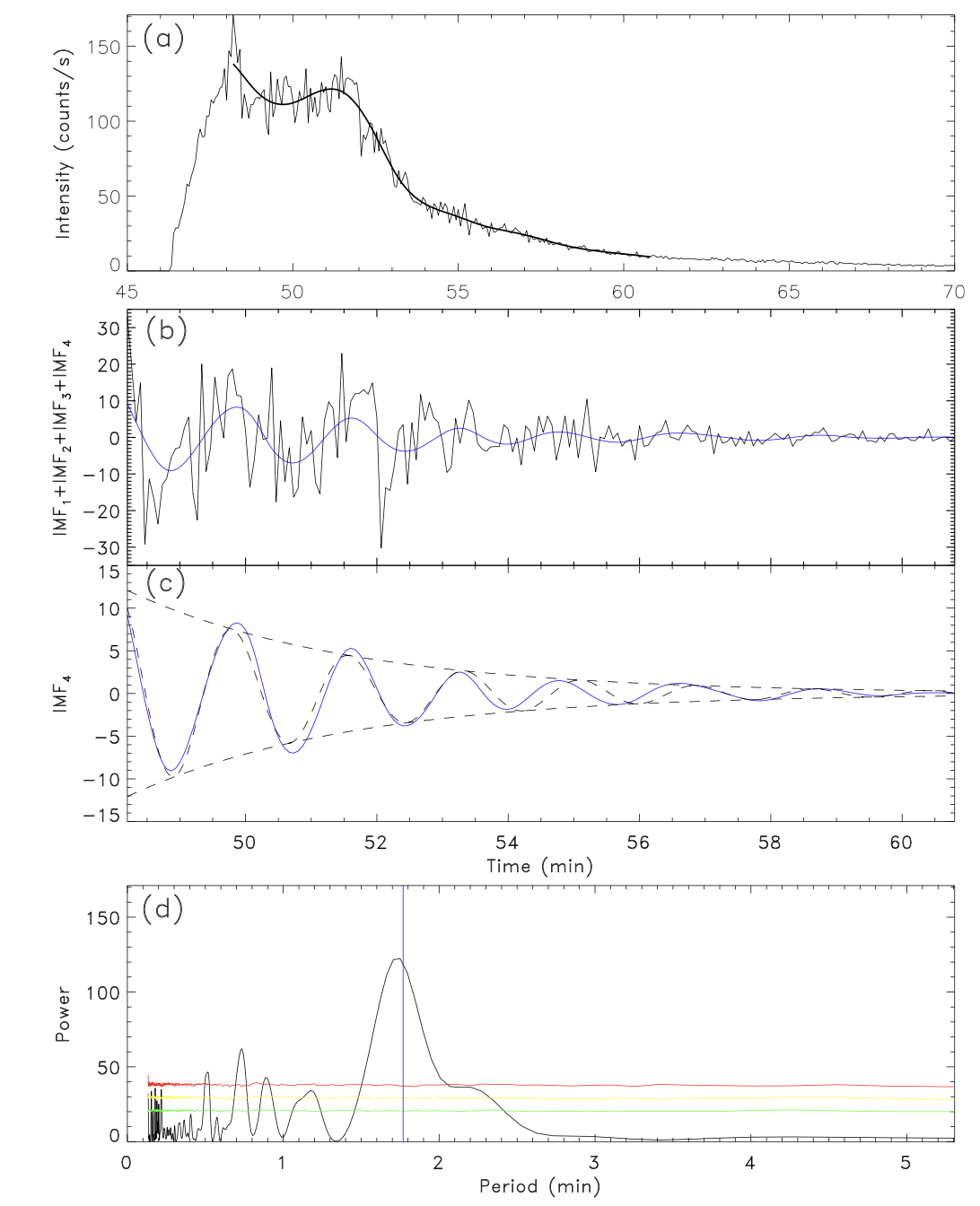}
\centering\includegraphics[width=2.5in]{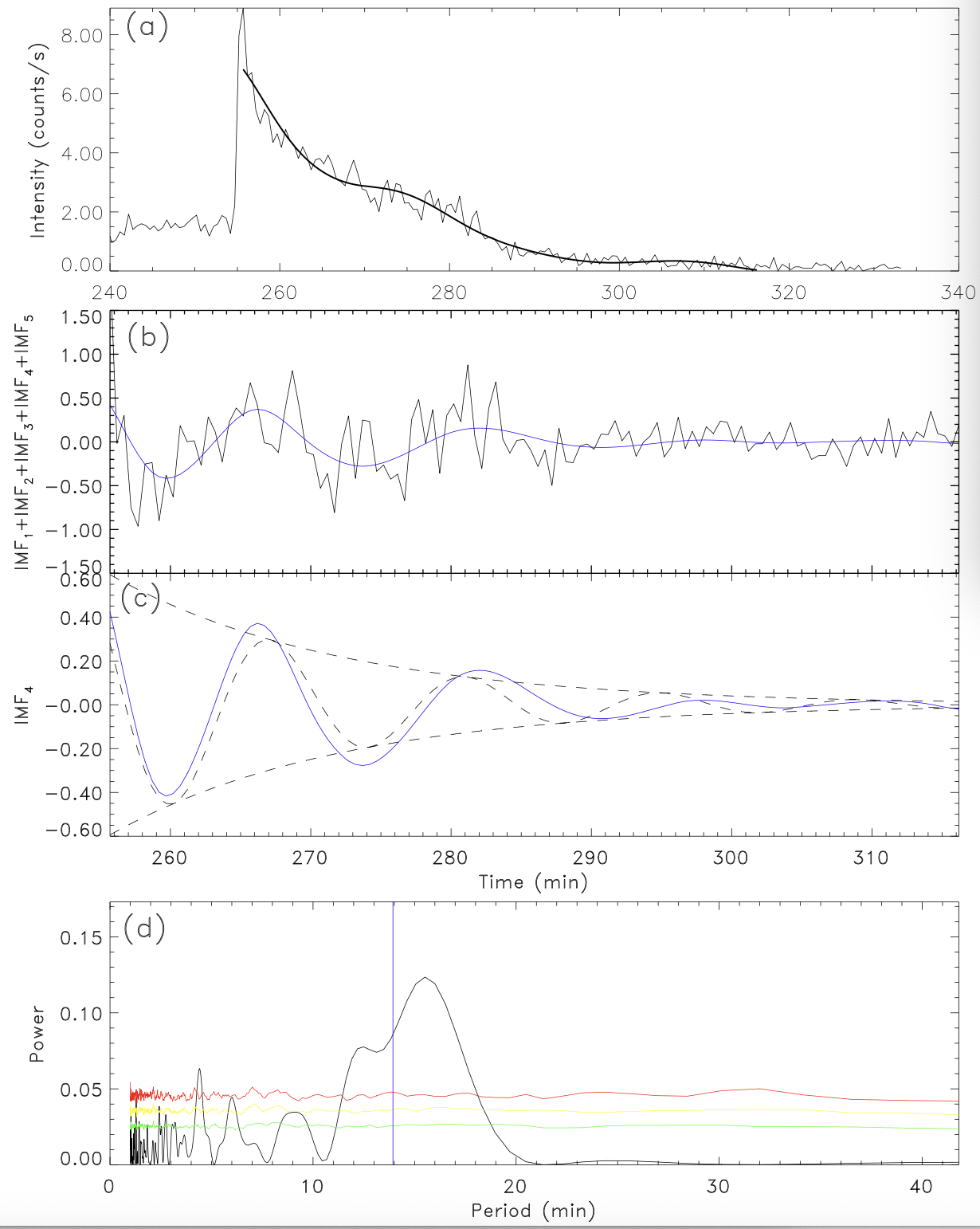}
\caption{Left: (a) X-ray light curve of a solar flare observed with RHESSI at the 3–24 keV energies, with  the trend obtained by
applying the EMD technique (thick solid line). (b) Residual signal obtained by detrending, and  damped oscillatory trend (blue line). (c) Best
fitting by a decaying harmonic oscillation ( black dashed curve). (d) Power spectrum of the residual signal given in panel (b). The blue vertical line is the
period obtained from fitting by the least-squares technique. The red, yellow, and green curves are 99\%, 95\%, and 90\% confidence intervals, respectively. Right: the same as left but an XMM-Newton light curve of a stellar flare (from \cite{Cho2016a}).}
\label{fig:cho0}
\end{figure}

Direct comparison of samples of QPPs in solar and stellar flares have been performed by \cite{Cho2016a}. In this study, attention was put also on the damping of the pulsations. This damping denotes a form of dissipation associated with the wave process that can be highly expected \cite{Nakariakov2000a}. 
The analysis involves flares observed in relatively comparable high-energy bands, with RHESSI \cite{Lin2002a} in the 3-25 keV range for the Sun and XMM-Newton \cite{Jansen2001a} in the range 0.3-2 keV for the stars (Fig.\ref{fig:cho0}). A point of strength of this work is that they use similar techniques to analyze QPPs both in the solar and stellar light curves, and they compare similar samples of 42 solar and 36 stellar flares with significant QPPs. They compare the periods and damping times. They find periods with a mean of about 1 min for QPPs in solar flares and about 15 min in stellar flares. The damping times scale proportionally to 1.5 and 30 min, respectively. 
The ratio of the damping times to the period is, therefore, about 1.5 for both solar and stellar flares. This scaling is quite robust because it is obtained in very similar energy bands, and suggests that probably there is a common mechanisms for producing QPPs in solar and stellar flare.

\begin{figure}[!h]
\centering\includegraphics[width=5in]{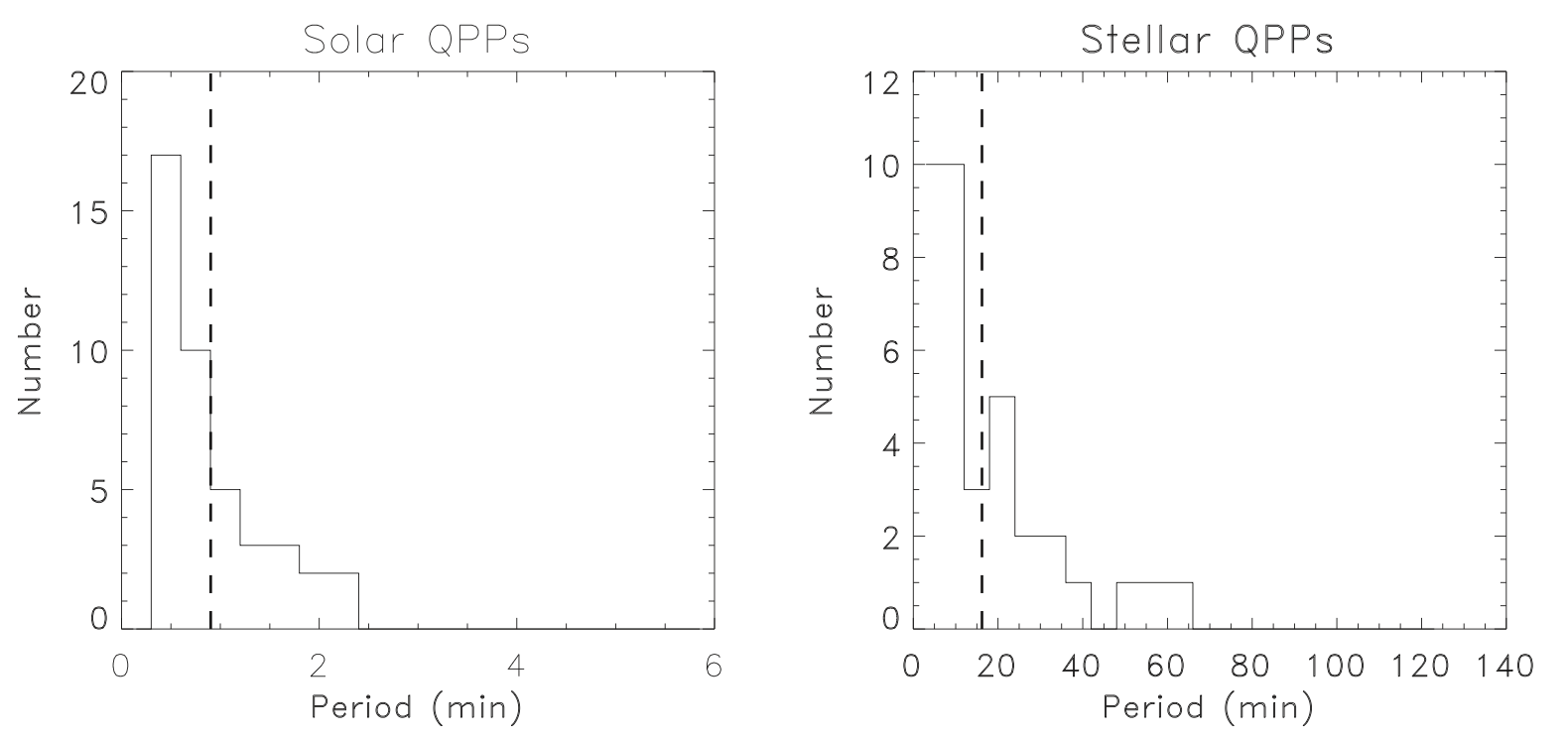}
\caption{Distributions of the periods for QPPs of solar (left) and stellar (right) flares, and  their mean values (vertical dashed lines).(from \cite{Cho2016a})}
\label{fig:cho1}
\end{figure}

As mentioned above, much larger statistical samples of hundreds QPPs are available from the long and continuous monitoring of GOES observations in the soft X-ray band \cite{Hayes2020a}, and from Kepler \cite{Pugh2016a} and TESS \cite{Ramsay2020a} surveys in the white light band. If we assume that they can be compared in spite of the different bands, although with quite broad distributions, {the core information from the statistical analysis is that the period of QPPs scales from about 1 min in solar flares (Section \ref{sec:qpp}\ref{sec:solar}) to 10-100 min (Section \ref{sec:qpp}\ref{sec:stellar}) in stellar flares (Fig.\ref{fig:cho1}). Together with the much larger energies involved in the stellar flares ($10^{33} - 10^{34}$ erg \cite{Guarcello2019a} than in solar flares ($< 10^{32}$ erg, e.g., \cite{Aulanier2013}),} this is a very simple and direct indication that stellar flares are generated in a framework of larger magnetic activity than on the Sun. On the one hand, then, {we have more intense magnetic fields (about 1 kG on average \cite{Wanderley2024} vs few G on the Sun \cite{Kotov2008}) that on scale} determine a larger energy release. This larger energy release make the QPPs in stellar flares well visible in the white light, whereas they are not on the Sun. On the other hand, these larger fields also imply spatially larger magnetic regions, and in particular longer magnetic loops, and presumably larger spacing between flare ribbons, {bright in the white light, that lead to the larger periods even in this band. }

\begin{figure}[!h]
\centering\includegraphics[width=2.5in]{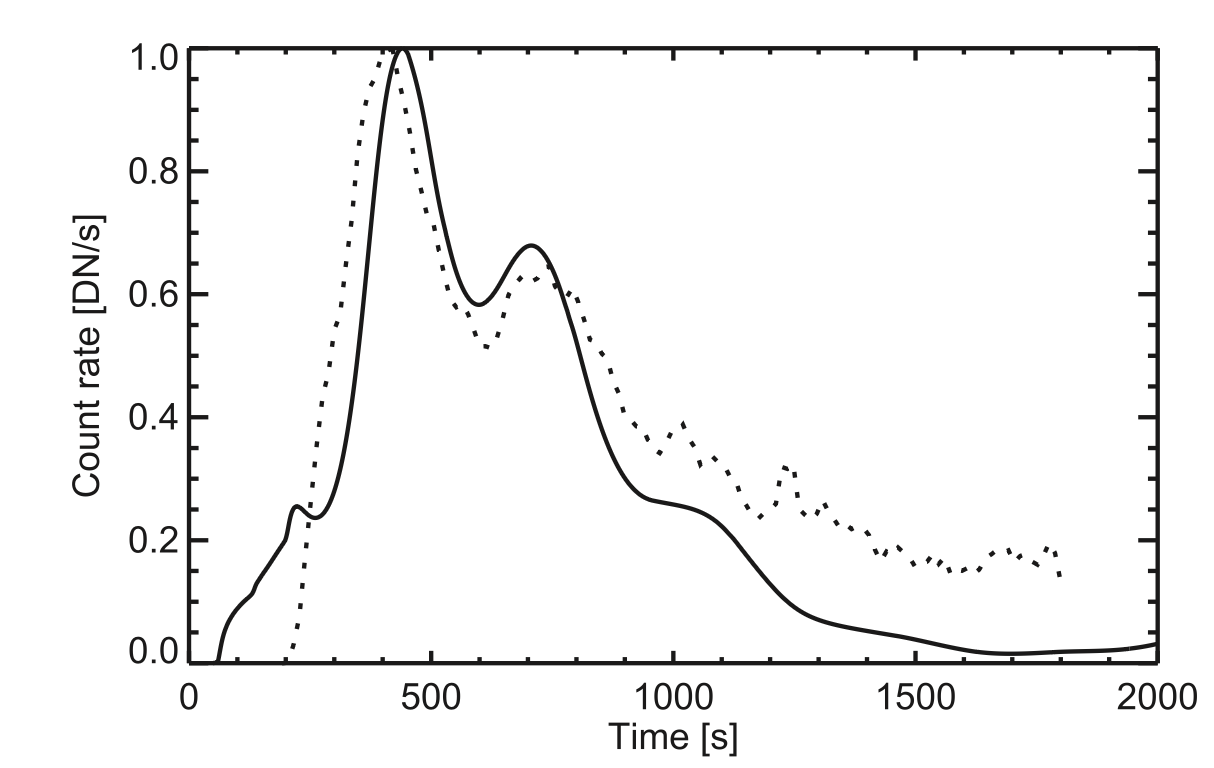}
\centering\includegraphics[width=2.5in]{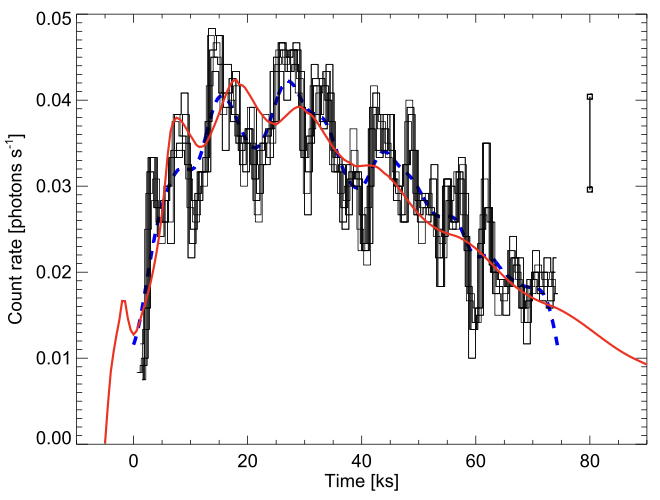}
\caption{Left: QPP in an EUV light curve of a solar microflare (dotted line) with the curve obtained by forward-modeling with a hydrodynamic loop simulation (solid line). Right: QPP in an X-ray light curve of a stellar flare,   with the curve obtained by the same forward-modeling with a hydrodynamic loop simulation (red line)(from \cite{Reale2018a,Reale2019a}).}
\label{fig:reale}
\end{figure}

In the hypothesis of a wave traveling along a tube, assuming a typical sound speed of 500 km/s (in coronal flaring conditions), a time scale of about 3000 s implies a tube length of a couple of solar radii, a really extended magnetic structure. This kind of result was found with detailed loop hydrodynamic modeling of two stellar flares observed in the Orion star forming region with the Chandra mission (COUP) which showed clear evidence of QPPs in the soft X-rays \cite{Reale2018a} (Fig.\ref{fig:reale}). In that case, the extremely high flare temperature and the long QPP periods of about 3 hrs led to infer a loop length even up to a few tens of solar radii. Although larger than the Sun, it is difficult that a young flaring star can entirely host such huge magnetic tubes, and therefore in this case one can even speculate that a magnetic channel linking the star to the accretion disk is flaring.

To date, this forward-modeling yields a light curve that can be directly compared with the observed one and thus, more than others, it actually provides a strong support to a propagating/standing hydromagnetic wave scenario, most probably slow magnetosonic waves in a proper fully MHD scenario \cite{Reale2019a}.  This is even more true if we consider that the same model could reproduce in detail also solar coronal loop microflare light curves \cite{Reale2019a} (Fig.\ref{fig:reale}). It is to be noted, that in this perspective, sausage MHD modes might also be compatible to reproduce the observation, which cannot be performed in the context of 1D hydrodynamic modeling.

Again, in the MHD scenario, only a strong single heating signal can trigger a coherent oscillation in a multi-stranded flux tube scenario. This means that QPPs and flares are intrinsically connected, and it is very unlikely to detect such oscillations in the framework of usual active region loop systems. Intermediate situations are possible as in microflares \cite{Reale2019a}, but also not \cite{Testa2020a}, thus supporting that the development of the oscillations might depend on the specific heating parameters. More specifically, in a general context, it was shown that the duration of the flare heat impulse is crucial to trigger pulsations in flare \cite{Reale2016a}. A short heat release and the following heat deficit are able to create a local pressure dip. The related pressure gradient then naturally travels back and forth along the flux tube, also in the form of a density wave. This determines local density excesses, non-linearly amplified in optically-thin emission, thus determining a density pulsating evolution. The short or long duration is referred to the sound traveling time along the flux tube, and therefore it changes with the tube length (and with the temperature). It is generally on the order of 1 min for the Sun \cite{Reale2019a}, but it can be even up to 1 hr on active stars \cite{Reale2018a}, thus indicating events on a very different scale, though possibly with similar mechanisms. This is absolutely general for both solar and stellar flares, so that the presence of pulsations of longer periods is again coherent with longer magnetic channels on stars than on the Sun, with a shared physical triggering mechanism. However, this does not mean that the details of the energy release are exactly the same on the Sun and the other more active stars, because pulsations are a by-product, not connected to the flare origin. For instance, energy release might be due either to a thermal front or to accelerated particles, but pulsations are triggered in both cases, and do not in general show signatures of the mechanism. In this case, QPPs cannot be used to investigate this aspect.
{New potentiality for quantitative comparison with observations comes from a recent flux rope model of flares which has been shown to generate restoring forces and therefore QPPs \cite{Soloviev2025}.}

{In the perspective of the wave nature of QPPs, it is also important to point out the possible presence of multiple periods within observed pulsations. Well-defined harmonics have been detected in a solar flare in the Nobeyama radio band and RHESSI hard X-ray band \cite{Inglis2009}. This is also the case with some strong stellar flares observed on Proxima Centauri and other low mass M stars in the XMM-Newton soft X-ray band  \cite{Srivastava_2013,Kolotkov_2021}, but also in the optical band  \cite{Doyle2022}. In all these cases the fundamental period is of a few tens of minutes, except for the last one, which goes down to a few minutes.}
{Though not properly surveyed systematically, it is worth mentioning some other observed features regarding QPPs in both solar and stellar flares. Mechanisms different from MHD waves seem to drive interesting pre-flare pulsations with increasing amplitude which have been detected with GOES on the Sun with quite long periods (8 - 30 min) \cite{Tan_2016}. These have been interpreted in terms of LRC current oscillations, similar to those found in tokamacs. A similar train of pulsations, but with random amplitude and in the optical band, has been detected also in  stellar flare observations with the Kepler mission \cite{Kuznetsov_2021}. }

\section{Summary and conclusions}


{Quasi-periodic pulsations (QPPs) observed in solar and stellar flares are reviewed and compared here, emphasizing results from large statistical surveys and homogeneous analyses. 
For solar flares, extensive studies using GOES X-ray data show that QPP periods are typically tens of seconds, with a log-normal distribution centered around 20 – 25 s. The QPP period correlates with flare duration and, wherever present, with flare ribbon separation, thus suggesting a dependence on magnetic loop length, and no correlation with flare intensity. Detection rates increase with flare class, and QPPs often exhibit damping. Multiple detection techniques are used (Fourier, wavelets, autocorrelation, machine learning). Several physical mechanisms have been proposed,  most notably MHD wave modes and oscillatory magnetic reconnection. }

{For stellar flares, QPPs have been sampled extensively in the white light thanks to the surveys by the Kepler and TESS missions, but X-ray observations are available also in the soft X-rays from XMM-Newton and Chandra missions. Periods are much longer, typically 10–100 minutes, and in some cases even hours. Large surveys find QPPs in only a minority of stellar flares (a few percent), and importantly, there is no correlation between QPP period and global stellar parameters (radius, temperature, flare energy), implying that QPPs are governed by local loop properties rather than global stellar properties. Damping properties are similar to solar QPPs when analyzed in comparable bands, with the ratio of damping time to period being approximately constant across solar and stellar cases. }

{Comparative studies suggest that, although the timescales differ by roughly one to two orders of magnitude, the underlying physics is likely the same. The longer periods in stars can be interpreted as evidence of larger magnetic structures and longer loops, consistent with stronger magnetic activity. Forward modeling of hydrodynamic loop oscillations successfully reproduces both solar and stellar QPP light curves, supporting a common wave-based mechanism (e.g., slow magnetosonic oscillations). }

There is plenty of room for future developments. In particular, it will be very interesting to have more extensive surveys of QPPs in proper high energy bands. On the theoretical side, further extensive forward-modeling efforts, possibly including the magnetic field in a full MHD description, and resulting into  detailed comparison to observations, are required to assess different driving mechanisms on the same testing ground.

\ack{The author acknowledges support from italian Ministero dell'Università e della ricerca (MUR).}


\bibliography{QPP_Reale_bib}
\bibliographystyle{RS}

\end{document}